\documentclass[
amsmath,amssymb,
pra,
preprint,
aps,
superscriptaddress,
nobibnotes
]{revtex4-1}

\usepackage{fontspec}
\usepackage{graphicx}
\usepackage{dcolumn}
\usepackage{bm}
\usepackage[colorlinks=true, allcolors=blue,breaklinks]{hyperref}
\usepackage{braket}
\usepackage{bbding}
\usepackage{natbib}
\usepackage{gensymb}
\usepackage{balance}
\usepackage{pifont}
\usepackage{longtable}
\usepackage{times}
\usepackage{microtype}

\makeatletter
\def\mcitePlusGraphical{\@gobble} 

\makeatother

\usepackage{newtxtext}
\usepackage{newtxmath}

\begin{document}

\title{Light-Driven Ferroic Switching Enables Reversible Control of Hydrogen Adsorption Thermodynamics}

\author{Xueqing Wan}
\affiliation{%
 Ministry of Education Key Laboratory for Nonequilibrium Synthesis and Modulation of Condensed Matter, Shaanxi Province Key Laboratory of Advanced Functional Materials and Mesoscopic Physics, School of Physics, Xi'an Jiaotong University, Xi'an 710049, China 
}%

\author{Zhenlong Zhang}
\affiliation{%
Ministry of Education Key Laboratory for Nonequilibrium Synthesis and Modulation of Condensed Matter, Shaanxi Province Key Laboratory of Advanced Functional Materials and Mesoscopic Physics, School of Physics, Xi'an Jiaotong University, Xi'an 710049, China 
}%

\author{Charles Paillard}
\affiliation{%
Smart Ferroic Materials Center, Physics Department and Institute for Nanoscience and Engineering, University of Arkansas, Fayetteville, Arkansas 72701, USA 
}%
\affiliation{%
Université Paris-Saclay, CentraleSupélec, CNRS, Laboratoire SPMS, 91190, Gif-sur-Yvette, France 
}%

\author{Jian Zhou}
\affiliation{%
Center for Alloy Innovation and Design, State Key Laboratory for Mechanical Behavior of Materials, Xi’an Jiaotong University, Xi’an 710049, China
}%

\author{Jinyang Ni}
\email{jyni@xjtu.edu.cn}
\affiliation{%
Ministry of Education Key Laboratory for Nonequilibrium Synthesis and Modulation of Condensed Matter, Shaanxi Province Key Laboratory of Advanced Functional Materials and Mesoscopic Physics, School of Physics, Xi'an Jiaotong University, Xi'an 710049, China 
}%

\author{Chuanlu Yang}
\email{ycl@ldu.edu.cn}
\affiliation{%
School of Physics and Optoelectronic Engineering, Ludong University, Yantai 264025, China
}%
\affiliation{%
Xinjiang Astronomical Observatory, Chinese Academy of Sciences, Urumqi 830011, China
}%

\author{Zhijun Jiang}
\email{zjjiang@xjtu.edu.cn}
\affiliation{%
 Ministry of Education Key Laboratory for Nonequilibrium Synthesis and Modulation of Condensed Matter, Shaanxi Province Key Laboratory of Advanced Functional Materials and Mesoscopic Physics, School of Physics, Xi'an Jiaotong University, Xi'an 710049, China 
}%

\author{Laurent Bellaiche}
\affiliation{%
Smart Ferroic Materials Center, Physics Department and Institute for Nanoscience and Engineering, University of Arkansas, Fayetteville, Arkansas 72701, USA 
}%
\affiliation{%
Department of Materials Science and Engineering, Tel Aviv University, Ramat Aviv, Tel Aviv 6997801, Israel
}%



\begin{abstract}
Reversible ultrafast switching of surface thermodynamics is highly desirable for hydrogen storage and catalysis yet remains elusive at the nanoscale. Here we demonstrate that photoinduced ferroic-order switching in two-dimensional ionic ferroelectric monolayers enables rapid, reversible control of hydrogen binding. In TiGeSe$_3$, carrier-density-driven redistribution of transition-metal 3\textit{d} orbital occupations triggers a sequential evolution from the ferroelectric ground state to paraelectric phases with staggered or Zig-Zag antiferromagnetic order. This switch continuously tunes the hydrogen adsorption free energy from 0.33 to 1.11 eV, shifting the interface from near-thermoneutrality to spontaneous desorption. Nonadiabatic dynamics indicate that electron-phonon coupling promotes nonthermal H release, while picosecond carrier recombination rapidly restores the initial ferroic order, closing an ultrafast reversible cycle. Generality is further validated in AgBiP$_2$Se$_6$ and CuInP$_2$S$_6$, establishing ferroic order as an optically addressable knob for dynamic thermodynamic reconfiguration beyond static design.
\end{abstract}

\maketitle	

As the fuel with the highest gravimetric energy density, hydrogen is poised to be a vital component of our future of renewable energy \cite{xu2024electrochemical,hassan2021monetization,majid2020renewable}. Electrochemical hydrogen storage has emerged as a promising solution, recognized for its high capacity, safety, and cycling stability \cite{yadav2022electrochemical,eftekhari2017electrochemical}. Despite this promise, the implementation of state-of-the-art materials—ranging from rare-earth to Ti-based alloys—is severely hindered by three fundamental constraints: excessively high operating temperatures, slow kinetics, and unsatisfactory cycle life \cite{xu2024electrochemical}. This disparity between technological potential and material limitations highlights a central challenge: achieving precise, reversible control of hydrogen storage without relying on energy-intensive thermal cycles or cumbersome pressure swings. 

Non-equilibrium control of material phases represents a critical frontier in condensed matter physics, providing a promising path to tune atomic-scale properties that could address pressing challenges for efficient hydrogen storage. Manipulation via external fields, such as optical and electric stimuli, offers a robust platform for achieving precise phase transition control. Specifically, electric fields can modulate hydrogen adsorption via phase transitions in polar materials \cite{zhou2010electric,liu2009electric,han2024insights}. This response is typically slow and often compromised by interfacial degradation. In contrast, femtosecond laser excitation enables structural switching on the subpicosecond scale, translating to a speed six orders of magnitude faster than conventional electric gating \cite{zhang2019electric, Zhang2019light,wang2017suppressing}.

Light-induced phase transitions have been shown to give rise to a suite of emergent phenomena, including photoinduced superconductivity \cite{ma2021large,lin2014critical}, topological states \cite{Bo2024}, reduced lattice thermal conductivity \cite{cazorla2024light}, and ultrafast electronic switching \cite{sternbach2021nanotextured,johnson2023ultrafast}. A primary advantage of utilizing light fields lies in their electrode-free operation, which circumvents the performance degradation mechanisms typically inherent to electrically driven architectures \cite{luchtefeld2023operando,xue20193d}. Additionally, light fields are typically less susceptible to lattice damage, ensuring slow degradation and long-term cycles. These characteristics are crucial for reversible processes like hydrogen storage, where the inherently transient states created by photoexcitation enable rapid cycling between adsorption and desorption configurations. At the microscopic scale, photogenerated nonthermal carriers can directly couple energy into chemical bonds, thereby facilitating hydrogen dissociation and enabling reversible storage \cite{brongersma2015plasmon,sandor2024ultrafast,avdizhiyan2023ultrafast,bakalis2024momentum,shen2025interplay}. Collectively, these attributes establish light field modulation as a robust and versatile paradigm for hydrogen storage. However, the fundamental mechanistic links bridging ultrafast photoinduced phase transitions to their resultant impact on chemical reactivity—specifically the thermodynamic landscape of hydrogen adsorption—remain largely elusive.

\begin{figure*}
\includegraphics[width=1.0\linewidth]{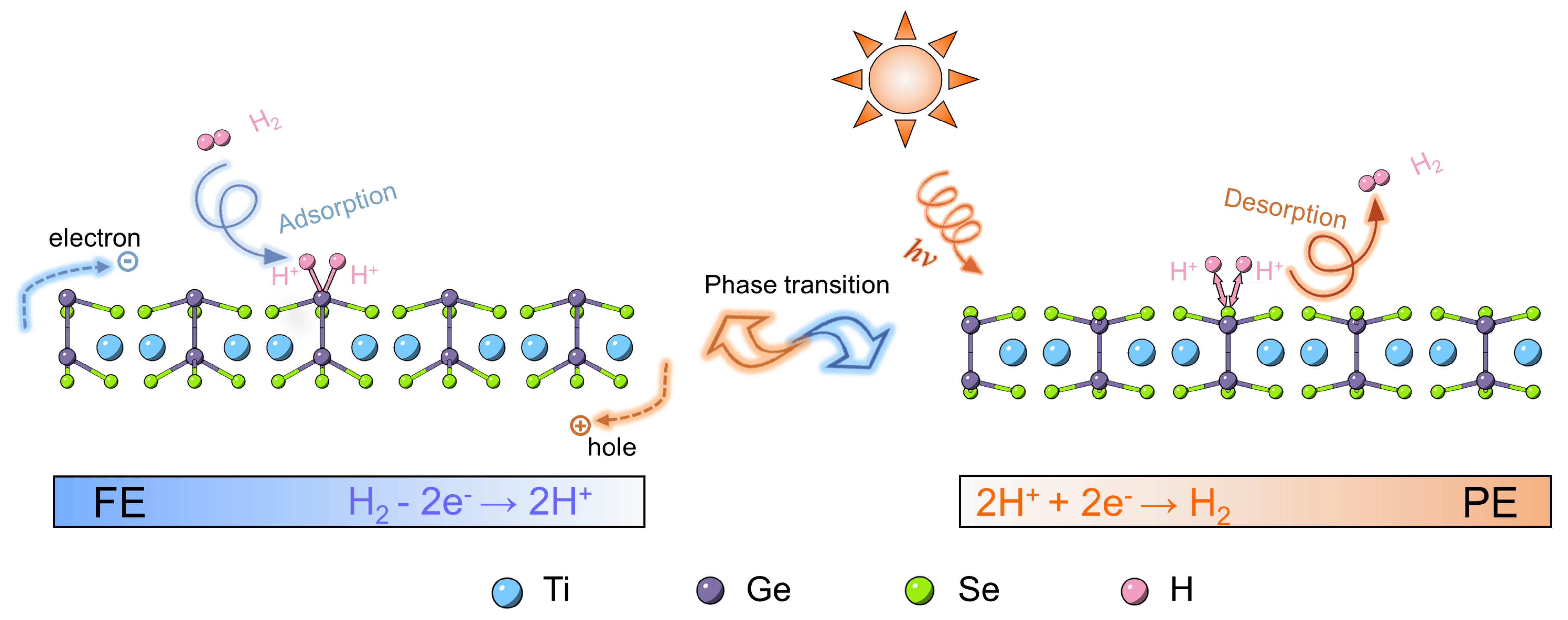}
\caption{Illustration of the underlying mechanism for photo-controlled hydrogen storage in the ionic ferroelectric of TiGeSe$_3$ monolayer. The left panel illustrates carrier separation driven by the intrinsic electric field in the FE phase without light. In contrast, the right panel shows accelerated energy transfer via strong electron-phonon coupling in the PE phase under light illumination.}
\label{fig1}
\end{figure*}

To bridge this fundamental gap, we report a highly efficient and reversible regulatory strategy based on light-induced ferroic switching in ionic ferroelectrics. Specifically, taking monolayer TiGeSe$_3$  \cite{wang2023alterferroicity,zhang2025landau} as a prototypical system, we demonstrate that ultrafast photoexcitation triggers a structural transition between ferroelectric (FE) and paraelectric (PE) phases. This transformation effectively shifts the Gibbs free energy change for hydrogen adsorption ($\Delta G_{\text{*H}}$) toward more positive values, providing the essential thermodynamic driving force for hydrogen desorption, as illustrated in Figure \ref{fig1}.

By combining $\Delta$  self-consistent field ($\Delta$SCF) excited-state calculations \cite{paillard2019photoinduced,paillard2023light} with nonadiabatic molecular dynamics (NAMD) simulations \cite{zheng2017phonon,zheng2019ab,chu2020low,yang2019ultrafast}, we reveal that laser excitation modifies the occupancy of conduction-band \textit{d} orbitals, destabilizing the ferroelectric distortion and triggering the phase transition. Notably, this photoinduced electronic reconstruction enhances both electron-phonon coupling and nonadiabatic coupling (NAC) in a light-intensity-dependent manner. These effects collectively promote nonthermal electron-driven bond weakening while ensuring rapid recovery of the ferroic order through electron-hole (\textit{e}-\textit{h}) recombination. Importantly, we establish the generality of this mechanism by validating it in monolayer CuInP$_2$S$_6$ and AgBiP$_2$Se$_6$. In these systems, the light-induced modulation of $\Delta G_{\text{*H}}$ significantly exceeds that achievable through conventional stimuli, such as epitaxial strain, external electric fields, or chemical doping. By tuning the photoexcited carrier density (\textit{n}$_\text{ph}$), $\Delta G_{\text{*H}}$ can be continuously driven across the thermoneutral limit, enabling reversible switching between hydrogen adsorption and desorption states in AgBiP$_2$Se$_6$. Together, these results establish light-driven ferroic order switching as a general design principle for ultrafast and reversible control of catalytic thermodynamics.

\begin{figure*}
\includegraphics[width=1.0\linewidth]{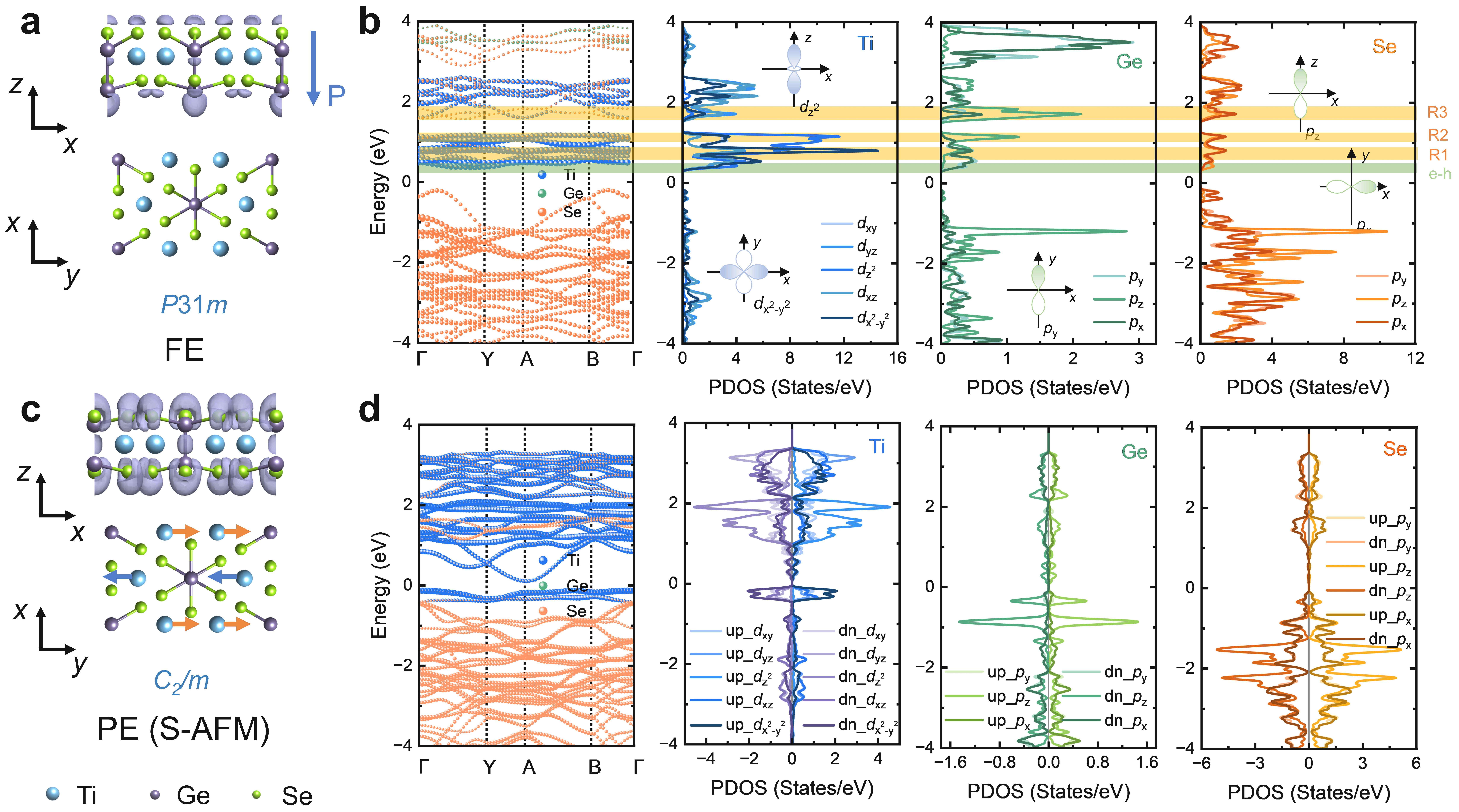}
\caption{Electronic properties of TiGeSe$_3$ monolayer.  Local charge densities, geometric structures, band structures, and PDOS of the (a, b) FE and (c, d) PE (S-AFM) phases in the TiGeSe$_\text{3}$ monolayer, respectively. The labels ``up'' and ``dn'' denote spin-up and spin-down channels, respectively. The arrows in (a, c) represent the directions of polarization and magnetic moment, respectively.}
\label{fig2}
\end{figure*}

We first consider the monolayer TiGeSe$_3$ as an example to illustrate the light-induced hydrogen storage mechanism by ferroic phase transition. TiGeSe$_3$ hosts two competing stable states with strikingly different electronic characteristics (Figure~\ref{fig2}). In the nonmagnetic FE phase (\textit{P}31\textit{m}), Ge-Ge dimer displacements induce an out-of-plane polarization of 3.88 pC/m via asymmetric Ge electron localization. This value is in close agreement with the 5.30 pC/m reported in a previous study \cite{wang2023alterferroicity} using the Berry phase method \cite{king1993}. The PE phase (\textit{C}$_{2}$/\textit{m}) without polarization is characterized by a staggered antiferromagnetic (S-AFM) order on neighboring Ti site \cite{wang2023alterferroicity,zhang2025landau}, labeled as PE\,(S-AFM). Interestingly, local charge density analysis reveals asymmetric electron localization near Ge atoms on the top in the FE phase, while the PE (S-AFM) phase shows the symmetrical characteristics and exhibits alternating magnetic moment direction between neighboring Ti sites (Figure~\ref{fig2}a and c).

The spontaneous polarization in the FE phase originates from the lone-pair activity of Ge dimers, which simultaneously drives a Ti valence change from Ti$^{4+}$ to Ti$^{3+}$ \cite{zhang2025landau}. This charge redistribution triggers a profound reorganization of Ti 3$d$ orbitals. While the PE (S-AFM) state maintains a 3$d^1$ configuration with occupied states in the valence band giving rise to magnetism, the FE state adopts an empty 3$d^0$ occupancy, characterized by a nonmagnetic nature. This switching of orbital order not only governs the ferroic competition but also fundamentally reshapes the electronic band structure. Specifically, the transition from the FE phase to the PE (S-AFM) phase undergoes an indirect-to-direct bandgap transition, accompanied by a gap reduction from 0.56 to 0.23 eV (Figure~\ref{fig2}b and d). These findings reveal a synergistic mechanism in which ferroic order and 3$d$-orbital occupancy are intrinsically coupled. Critically, this coupling dictates a deterministic pathway for optical control: photoexcited electrons directly populate the unoccupied Ti 3$d$-orbitals, thereby destabilizing the FE state in favor of the PE (S-AFM) configuration. Consequently, TiGeSe$_3$ emerges as a prototypical two-dimensional (2D) ionic ferroelectric in which light-driven structural switching is dictated by orbital-selective excitations, establishing a rigorous microscopic foundation for the reversible regulation of hydrogen storage.

\begin{figure}
\includegraphics[scale=0.06]{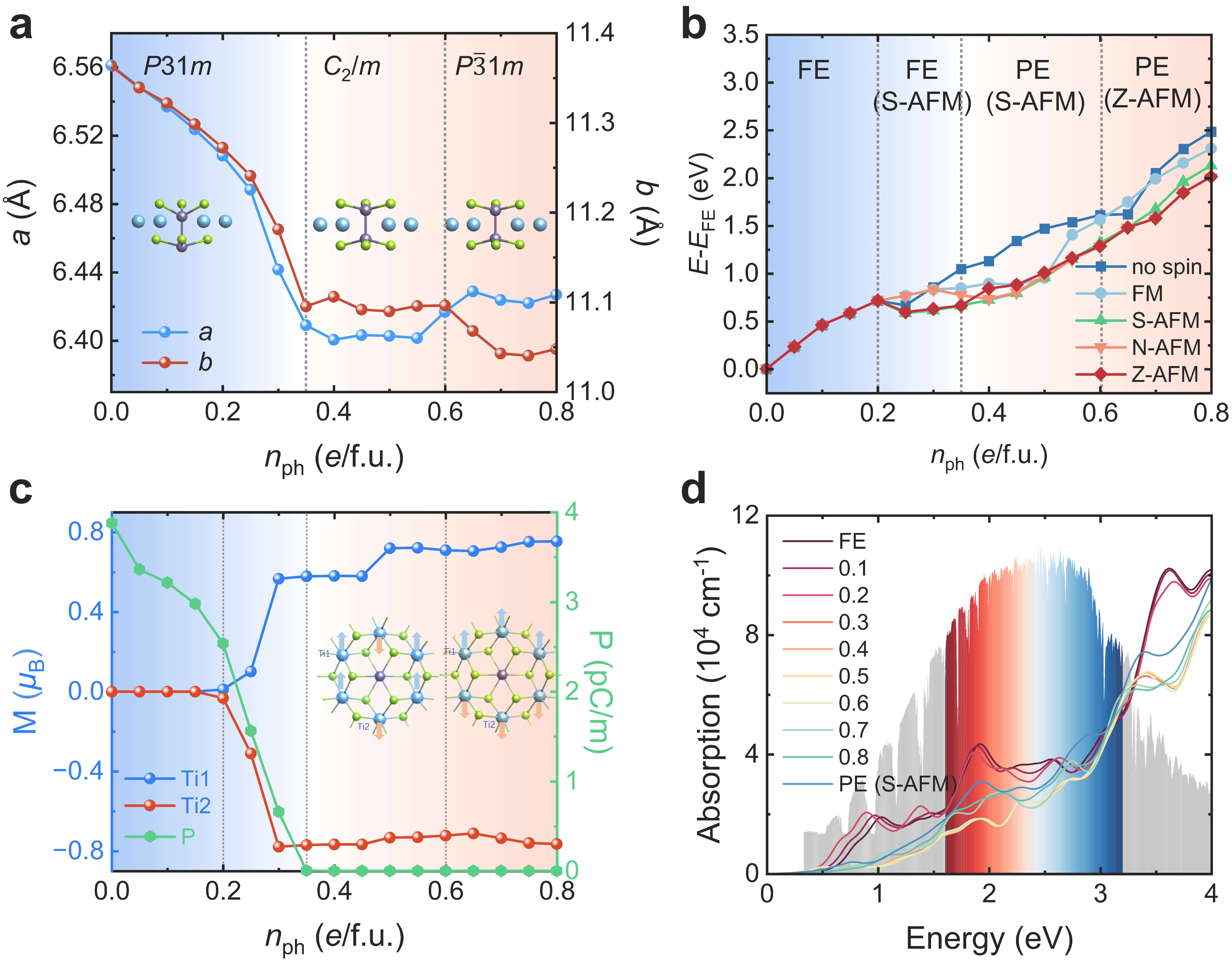}
\caption{Changes in geometry and electronic structures of TiGeSe$_3$ under light. (a) In-plane lattice constants versus $n_\text{ph}$. (b) Relative energy of various spin configurations—including ferromagnetic (FM), S-AFM, Néel-type antiferromagnetic (N-AFM), and Z-AFM referenced to the FE ground state devoid of phototexcited $e$-$h$ pairs. Calculations neglecting spin polarization (no spin) are also included. (c) Magnetic moments of two Ti atom types and polarization under different \textit{n}$_\text{ph}$. The blue and orange areas represent the FE and PE phases, respectively. (d) The effective optical absorption coefficients for the FE and PE (S-AFM) states in the absence of illumination, alongside the FE state under different \textit{n}$_\text{ph}$. The background represents the solar irradiance spectrum, provided as a visual reference for comparison.}
\label{fig3}
\end{figure}

The reduction of bandgap from FE to PE (S-AFM) phase indicates that optical dressing promotes the system to a higher total energy state, especially when photon energy is well-above the bandgap. As illustrated in  Figure \ref{fig3}a-c, photoexcitation induces a significant reorganization of the structural and electronic properties in TiGeSe$_3$. The FE order is progressively suppressed with increasing $n_\text{ph}$, as evidenced by the monotonic reduction of the polarization magnitude. At low $n_\text{ph}$ ($<$ 0.2 \textit{e}/f.u.), the system is characterized by the persistence of the FE state without spin polarization, which is consistent with the previous work \cite{wang2023alterferroicity,zhang2025landau}. For $n_\text{ph}$ ranging between 0.2 and 0.35 \textit{e}/f.u., a distinctive intermediate regime emerges, characterized by the coexistence of persistent ferroelectric polarization and emergent antiparallel magnetic moments on adjacent Ti sites; this state is identified as the FE (S-AFM) phase. Beyond a critical threshold of 0.35 \textit{e}/f.u., the polarization undergoes complete quenching, marking the transition to a dominant PE (S-AFM) phase. Interestingly, as \textit{n}$_\text{ph}$ increases, the magnetic order of the PE phase undergoes a transition from S-AFM to Zig-Zag AFM (Z-AFM) ordering. This magnetic evolution originates from the fundamental redistribution of Ti 3$d$ orbitals, as evidenced by the projected density of states (PDOS) in  Figure S2. As electrons are excited, the electrons progressively occupy the initial empty $3d^0$ states, effectively reducing Ti$^{4+}$ to Ti$^{3+}$. This orbital filling suppresses the Ge-Ge dimer distortion responsible for ferroelectricity while stabilizing the AFM order with PE phase.

Similarly, the FE-to-PE transition is accompanied by a concomitant reconfiguration of the electronic structure, characterized by an indirect-to-direct bandgap transformation and the emergence of prominent 3$d$-state density near the valence band maximum (VBM) ( Figure S2). TiGeSe$_3$ further distinguishes itself by its strong light-matter coupling, as shown in  Figure~\ref{fig3}d. The significant light absorption peaks located in the visible light region correspond to the large density of states region represented by R1, R2, and R3 in   Figure \ref{fig2}b. Notably, all phases exhibit pronounced absorption peaks in the visible range, with coefficients up to 8$\times10^4$ cm$^{-1}$. This performance is on par with or exceeds that of representative 2D semiconductors, such as g-C$_3$N$_4$ \cite{yao2022density}, InSe \cite{sun2022high}, and PtI$_2$ \cite{ge2023two}. Importantly, the fluence required to reach 0.8 \textit{e}/f.u. is only 21.93 mJ/cm$^2$ (calculated by eq~S18), which can be achieved by solar energy \cite{gueymard2002proposed}. The calculated solar energy utilization efficiencies (3.13--3.83\%) surpass those of several established semiconductors, including Ga$_2$Se$_3$, In$_2$Se$_3$, and AgInP$_2$S$_6$ \cite{fu2018intrinsic,wan2023boost}. Collectively, these findings establish TiGeSe$_3$ as a uniquely photosensitive FE monolayer, where light not only triggers FE-to-PE switching but also unlocks an exotic coexistence regime of polarization and magnetism, offering a deterministic microscopic pathway for photon-controlled hydrogen storage.

\begin{figure*}
\includegraphics[scale=0.05]{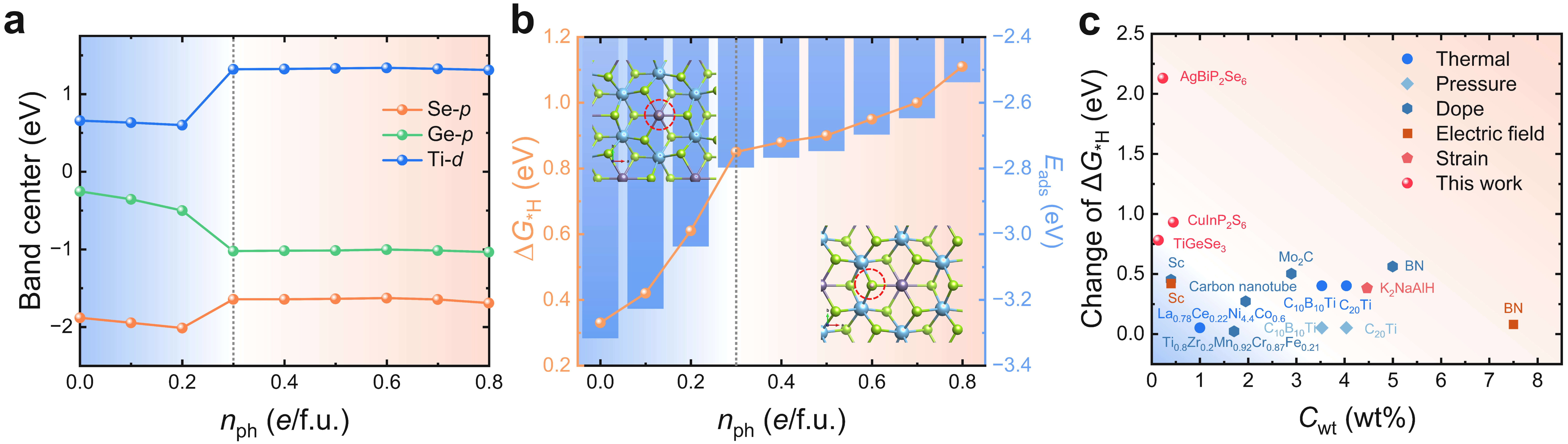}
\caption{Changes in band centers and $\Delta G_{\text{*H}}$s of TiGeSe$_3$ under light and the comparison with other reported materials. (a) Variation in the $d$/$p$-band centers of Ti, Ge, and Se atoms in the TiGeSe$_3$ monolayer with different \textit{n}$_\text{ph}$. (b) $E_{\text{ads}}$ and $\Delta G_{\text{*H}}$ for hydrogen storage of an H atom adsorbed on the Ge or Se sites as a function of \textit{n}$_\text{ph}$. Calculations performed at 0.1 \textit{e}/f.u. increments identify 0.3 \textit{e}/f.u. as the critical excitation threshold that governs the concomitant transitions in both the band center and $\Delta G_{\text{*H}}$. (c) A comparison of the $\Delta G_{\text{*H}}$ regulation capability in TiGeSe$_3$, AgBiP$_2$Se$_6$, and CuInP$_2$S$_6$ with other widely used regulatory measures for hydrogen storage materials \cite{parkar2024hydrogen,verma2024study,qiao2023effects,gao2025rationally,seif2016new,shi2017phosphorus,xia2021high,baaddi2024effect,li2020single,zhou2017improvement,zhou2010electric}}.
\label{fig4}
\end{figure*}

The binding strength between hydrogen and slab is closely tied to the position of the $d$- or $p$-band center relative to the Fermi level, a well-established descriptor for catalytic and storage performance \cite{zhu2021intrinsic,jiao2022descriptors,norskov2011density,zhang2023engineering}. In TiGeSe$_3$, the FE-to-PE transition induces a substantial shift in these band centers (Figure \ref{fig4}a), coupled with the evolution of PDOS under light. As revealed in  Figure S2, the isolated band is mainly contributed by Ge atoms, implying a strong electron localization; its disappearance enhances charge delocalization, thereby facilitating hydrogen desorption \cite{zhang2025unraveling,yu2010conduction,irie2007ag+}. Concomitant with the structural distortion of Ge-Ge pairs, the preferred adsorption sites also shift: hydrogen binds to Ge atoms in the FE phase but relocates to Se atoms once the PE (S-AFM) order is established at 0.35 \textit{e}/f.u. ( Figure~\ref{fig4}b). In general, hydrogen atoms exhibit more robust adsorption on non-metal elements, primarily via the formation of strong polar covalent bonds. Consequently, as $n_{\text{ph}}$ increases from 0.0 to 0.35 $e$/f.u., the hydrogen adsorption energy on the slab ($E_{\text{ads}}$) shifts toward less negative values (from $-$3.31 to $-$2.57 eV), indicating that hydrogen atoms are more prone to desorption. This trend confirms that the FE phase exhibits an ideal affinity for hydrogen atoms. Importantly, the light-driven PE phase transition inherently facilitates the efficient hydrogen desorption.

From a thermodynamic perspective, the observed switching behavior aligns perfectly with Sabatier's principle \cite{sabatier1920catalyse, chen2024unusual}, which is excessively strong adsorption in the FE state that hinders hydrogen release. Therefore, photoinduced weakening of the adsorption significantly accelerates desorption kinetics. Therefore, we calculated the $\Delta G_{\text{*H}}$ in the hydrogen storage process by eqs~S12-S15. Remarkably, our strategy enables a giant modulation of the hydrogen adsorption energetics, with $\Delta G_{\text{*H}}$ shifting by as much as 0.78 eV (from 0.33 to 1.11 eV). As shown in  Figure \ref{fig4}c, this regulatory strategy is considerably larger compared to conventional methods, including thermal, pressure-induced, strain, dope, and electric field, underscoring the exceptional efficacy of light as a control parameter. The higher $\Delta G_{\text{*H}}$ under illumination signifies a weakened adsorption strength, which markedly promotes the desorption process. Although the gravimetric capacity ($C_\text{wt}$) of the TiGeSe$_3$ monolayer is modest (0.14~wt\%), comparable to that of other emerging materials such as Ni@Ti$_2$C (0.55~wt\%) \cite{kim2024interfacial} and pristine Mg-metal organic framework (MOF) (0.1~wt\%) \cite{liu2024mg}, its key breakthrough lies in achieving reversible ultrafast optical switching between strong adsorption and facile desorption, thereby opening a fundamentally new pathway for light-driven hydrogen storage.

\begin{figure}
\includegraphics[scale=0.052]{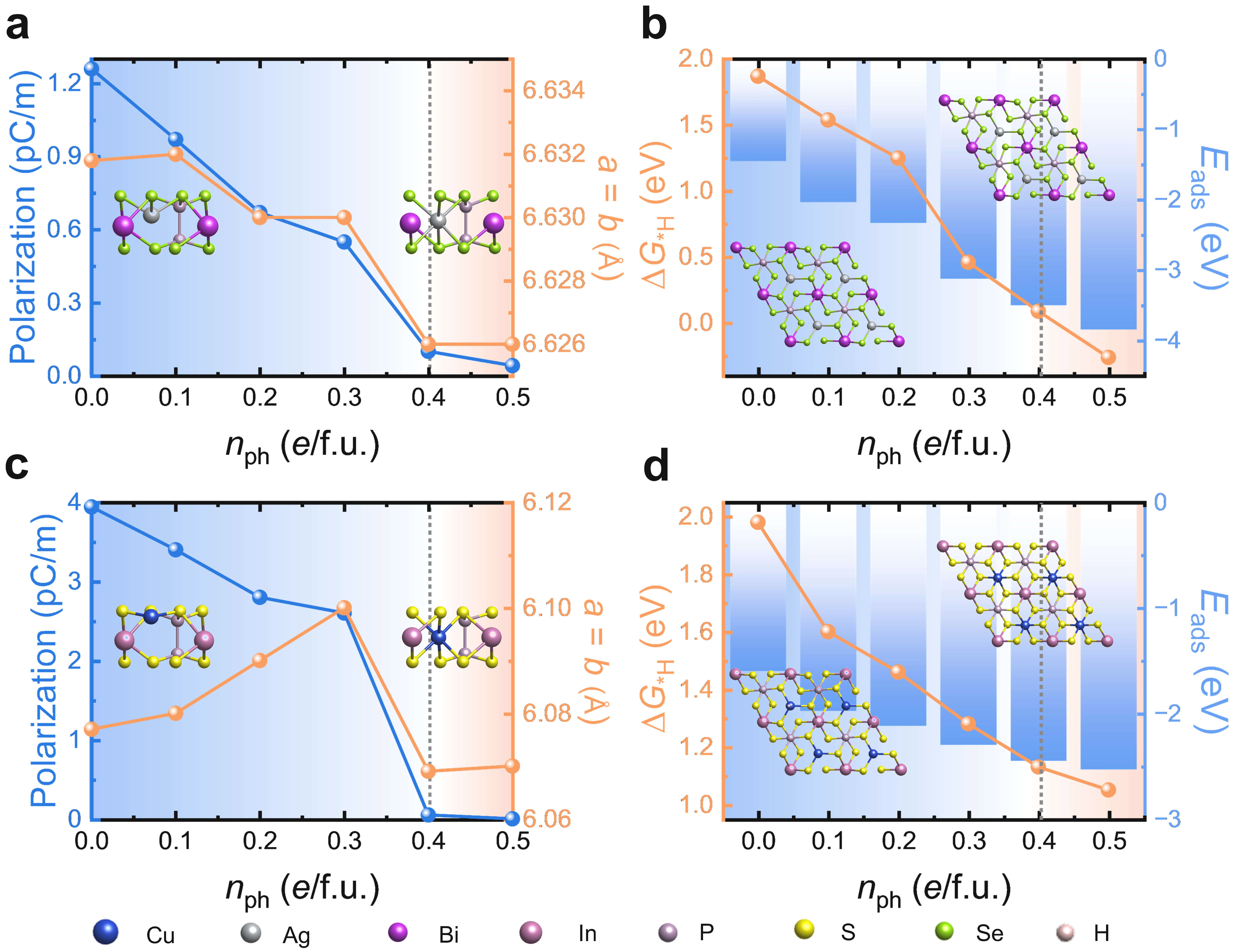}
\caption{Changes in geometric structure and hydrogen adsorption under light for layered metal phosphorus chalcogenides monolayers. Evolution of (a, c) polarization and in-plane lattice constants, and (b, d) $\Delta G_{\text{*H}}$ and $E_{\text{ads}}$ for hydrogen, as a function of \textit{n}$_\text{ph}$ in (a, b) AgBiP$_2$Se$_6$ and (c, d) CuInP$_2$S$_6$ monolayers, respectively. The insets in panels (a, c), and (b, d) depict the geometric configurations and corresponding hydrogen adsorption sites before and after the phase transition, respectively.}
\label{fig5}
\end{figure}

Additionally, the generality of this mechanism extends to layered metal phosphorus chalcogenides, providing experimental credibility for this regulatory approach \cite{sun2019strain, qi2018two}. Monolayers of FE AgBiP$_2$Se$_6$ and CuInP$_2$S$_6$ have been demonstrated as excellent photocatalysts \cite{liu2022excited, fan2021highly, ju2019tunable, xu2017monolayer}. As shown in Figure \ref{fig5}a and c, the out-of-plane polarization in these materials originates from the off-centering displacements of Ag/Cu, yielding polarization values of 1.26 pC/m and 3.94 pC/m, respectively \cite{liu2022excited, xu2017monolayer}. Our results show that both systems transition from the FE phase (space group \textit{P}3) to the PE phase (space group \textit{P}312) when \textit{n}$_\text{ph}$ reaches 0.4 $e$/f.u. Notably, the PE phases of AgBiP$_2$Se$_6$ and CuInP$_2$S$_6$ exhibit significantly enhanced hydrogen adsorption abilities, with adsorption energies of $-$0.26 eV and 1.05 eV at 0.5 $e$/f.u., respectively, compared to their FE counterparts (1.87 eV and 1.98 eV at 0.0 $e$/f.u.), as illustrated in Figure \ref{fig5}b and d. The corresponding $C_\text{wt}$ for eight H atoms adsorbed on 2$\times$2$\times$1 supercells of AgBiP$_2$Se$_6$ and CuInP$_2$S$_6$ are calculated to be 0.23\% and 0.45\%, respectively. Importantly, as the \textit{n}$_\text{ph}$ in AgBiP$_2$Se$_6$ is increased from 0.4 to 0.5 $e$/f.u., the $\Delta G_{^*\text{H}}$ exhibits a sign change, shifting from 0.09 to $-$0.26 eV. This change indicates a clear transition from energetically favorable desorption to favorable adsorption. Governed by ionic displacements during the phase transition, this mechanism represents a generalizable strategy applicable to a broad range of ionic ferroelectrics, analogous to the behavior discussed for the TiGeSe$_3$ monolayer. The key advantage of this approach lies in its ability to achieve precise, on-demand hydrogen release using simple light illumination, offering a more energy-efficient and controllable alternative to conventional thermally driven systems. These findings establish the ionic ferroelectric family as a compelling platform for next-generation hydrogen storage technologies, offering a versatile template for light-controlled energy solutions.

The key finding of this work is an experimentally feasible photocontrolled phase transition that drives a surface thermodynamic switching process. Among the studied materials, the CuInP$_2$S$_6$ system provides a particularly relevant experimental reference, as it has been widely applied in the field of photocatalysis \cite{lin2020ferroelectric,yu2021few} and achieves enhanced photocatalytic CO$_2$ reduction efficiency through electrically poled FE-PE transition \cite{chiang2024manipulating}. In addition, the laser-induced hydrogen desorption from a Ru(0001) surface has been observed in experiments \cite{Denzler2003,Fuchsel2012}. These studies provide experimental precedents and support the feasibility of probing the proposed correlation between ferroelectric phase transitions and hydrogen adsorption/desorption kinetics under light. Experimentally, ferroic order switching may be monitored by piezoresponse force microscopy (PFM) or photo-assisted PFM, transient carrier/lattice dynamics can be resolved by ultrafast pump-probe spectroscopy, and the evolution of surface H-binding states may be followed by in situ X-ray absorption spectroscopy and diffuse reflectance infrared Fourier transform spectroscopy. The amount and temperature dependence of desorbed hydrogen species can be further quantified by temperature-programmed desorption. Unlike conventional methods, our light-driven mechanism provides a non-contact, reversible, and ultrafast control route. In particular, the calculated $\Delta G_\text{*H}$ modulation reaches 0.78 eV in TiGeSe$_3$, which exceeds most values obtained by conventional regulation strategies summarized in Figure \ref{fig4}c. Conceptually, this work resembles femtosecond-laser-induced recombinative H$_2$ desorption from hydrogen-covered metal surfaces \cite{Denzler2003,Fuchsel2012,chiang2024manipulating} and charge redistribution during structural phase transitions, where nonthermal electronic excitation and charge delocalization can drive ultrafast desorption. However, our approach also relies on the reversible switching of ferroic order in ionic ferroelectrics, thereby reconfiguring the thermodynamics of hydrogen adsorption.

\begin{figure*}[h]
\includegraphics[scale=0.05]{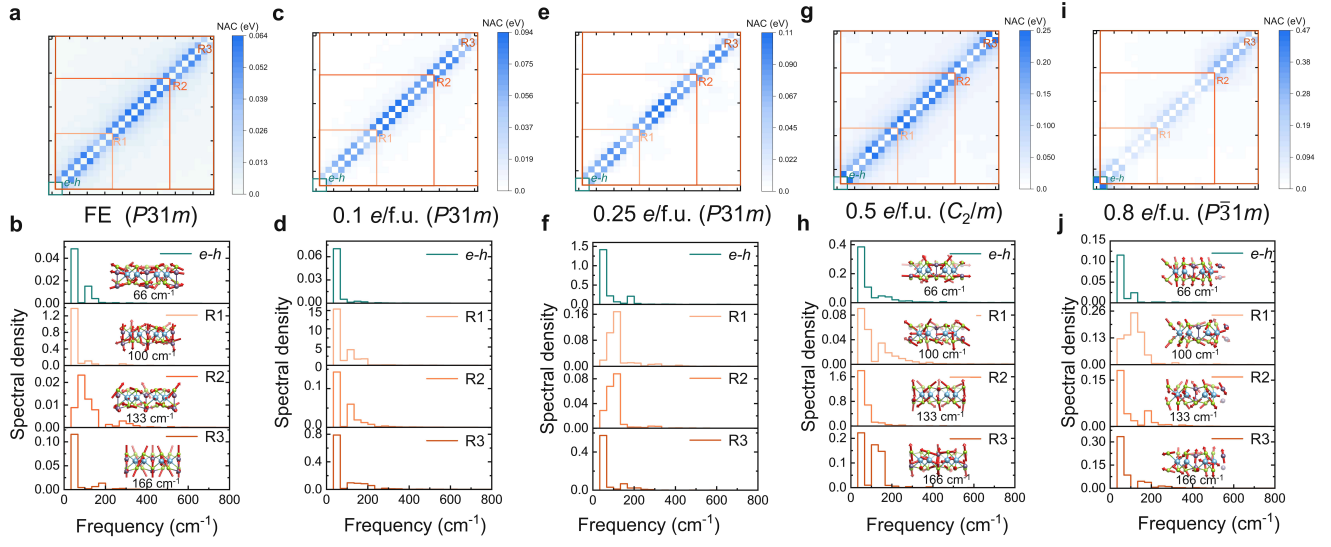}
\caption{Changes in NAC values and spectral densities of TiGeSe$_3$ under light. The averaged values of NAC between different states for the TiGeSe$_3$ monolayer with \textit{n}$_\text{ph}$ of (a) 0.0, (c) 0.1, (e) 0.25, (h) 0.5, and (i) 0.8 \textit{e}/f.u., respectively. The Fourier transforms for the normalized ACFs of the energy gap fluctuation between the acceptor and donor states in the R1, R2, R3, and $e$-$h$ recombination processes for the TiGeSe$_3$ monolayer with \textit{n}$_\text{ph}$ of (b) 0.0, (d) 0.1, (f) 0.25, (h) 0.5, and (j) 0.8 \textit{e}/f.u., respectively. The insets depict the characteristic phonon vibrational modes for the respective phases at various frequencies.}
\label{fig6}
\end{figure*}

To capture effects of the carrier transfer and recombination on the phase transition and hydrogen desorption, we performed NAMD simulations for the FE phase and its photoinduced counterparts at \textit{n}$_\text{ph}$ of 0.1, 0.25, 0.5, and 0.8 \textit{e}/f.u. (Figure~\ref{fig6}) in monolayer TiGeSe$_3$. To explicitly address the nonequilibrium carrier dynamics, including non-adiabatic effect in ${\Delta}$SCF framework, we performed the Hefei-NAMD code \cite{zheng2019ab}, in which ${\Delta}$SCF is integrated into the \textit{ab initio} molecular dynamics (AIMD), enabling real-time tracking of wave-function evolution under specified \textit{n}$_\text{ph}$. Here, the NAC factor is given by \cite{zheng2017phonon,zheng2019ab}:
\begin{equation}
d_{jk} \cdot \hat{R} = \langle \psi_j | \nabla_R | \psi_k \rangle \hat{R} = \frac{\langle \phi_j | \nabla_R | \phi_k \rangle}{\epsilon_j - \epsilon_k} \hat{R} 
\label{eq3}
\end{equation}

\noindent where \(\phi_j\) and \(\phi_k\) are the Kohn-Sham (KS) orbitals, \(\epsilon_j\) and \(\epsilon_k\) are the diabatic energies of states \(j\) and \(k\), and \(\hat{R}\) is the nuclear motion. The NAC factor is decided by the energy gap ($\epsilon_j - \epsilon_k$), the electron-phonon coupling strength ($\langle \phi_j | \nabla_R | \phi_k \rangle$), and nuclear velocity ($\hat{R}$). The band structures and KS orbitals of different configurations (Figures \ref{fig2}b, d, S2, and S6) display that the light excitation reduces the energy gaps associated with R1-R3 processes, where the R1, R2, and R3 processes correspond to the significant peaks in the visible light region (Figure \ref{fig3}d). The high-frequency and high-intensity vibration modes are attributed to the high nuclear velocity and strong electron-phonon coupling strength, respectively, as evidenced by the spectral density results. As illustrated in the insets of Figure \ref{fig6}b, h, and j, the vibration modes at various frequencies exhibit subtle variations in the TiGeSe$_3$ monolayer with different symmetries. The emergence of high-frequency peaks at 133 and 166 cm$^{-1}$ in the \textit{C}$_2$/\textit{m} and $P\bar{3}1m$ phases leads to the higher nuclear velocity relative to the FE phase of TiGeSe$_3$. The enhanced PDOS of the \textit{z}-direction orbital contributes to the higher intensity out-of-plane vibrational modes, which usually lead to stronger electron-phonon coupling due to the increased orbital overlap. The involvement of the unique vibrational modes in the R1 is primarily contributed by \textit{d}$_{x^2-y^2}$ of Ti in FE phase (Figure \ref{fig2}b), which couples with the in-plane vibration of Se and Ti atoms at 66 cm$^{-1}$. However, the contribution of this orbital gradually weakens, whereas that of the \textit{d}$_{z^2}$ orbital strengthens (coupled with the out-of-plane vibration of Ti atoms) under light (Figure S2). Moreover, the \textit{d}$_{z^2}$ orbital is the main contributor for the R2 process in all configurations (Figures \ref{fig2}b, d, and S2), which shows the electron-phonon interaction plays an important role in carrier relaxation. Regarding the R3, the \textit{p}$_z$ orbital of Ge tends to weaken, the \textit{d}$_{xz}$ and \textit{d}$_{yz}$ (the displacement of Ti atoms along the \textit{x}- and \textit{z}-directions) dominate this process, so the vibrations along \textit{x}, \textit{y}, and \textit{z} directions are strong (Figures \ref{fig2}b, d, and S2). Modifications in vibrational modes and the emergence of intensified peaks reflect a significantly enhanced electron-phonon coupling during the light-induced structural phase transition. These effects collectively amplify the NAC factor, thereby accelerating electron relaxation and energy transfer to the lattice.

As described in Section 3 of the Supporting Information, the stronger NAC and longer dephasing time will result in a faster carrier migration rate. The pure-dephasing time is obtained by optical response formalism \cite{long2016quantum}, through performing a Gaussian function $\exp(-0.5(t/\tau)^2)$ for the following equation:
\begin{equation}
D(t) = \exp\left[ -\frac{\Braket{(\delta U)^2}_T}{\hbar^2} \int_0^t d\tau_1 \int_0^{\tau_1} d\tau_2 \, C(\tau_2) \right]
\label{eq:dephasing}
\end{equation}
where $\delta U$ is the deviation of the energy gap from the average value and $C(t)$ is the normalized autocorrelation function (ACF) of the energy gap \cite{jaeger2012decoherence}. Therefore, the stronger correlation between the orbitals (Figure S6) leads to a longer dephasing time. Interestingly, the dephasing times in R1, R2, and R3 are longer than the $e$-$h$ recombination process, producing the rapid relaxation time under light as listed in Table S1. Indeed, the carrier lifetimes shrink dramatically with increasing photoexcitation from 0.66 to 0.08 ps (R1), 0.34 to 0.06 ps (R2), and 1.52 to 0.07 ps (R3). These faster electron transfer rates are attributed to the enhanced electron-phonon coupling strength, which demonstrates the stronger interaction between electrons and the lattice structures under light. Hence, the accelerated relaxation facilitates the transfer of energy from electrons to the slab, promoting hydrogen desorption in the light-induced PE (S-AFM) and PE (Z-AFM) phases. At the same time, $e$-$h$ recombination governs the recyclability of the system, as electrons and holes vanish from the conduction and valence bands. With the increase of $n_\text{ph}$, the contributions of the conduction band minimums (CBMs) and VBMs are transforms from the \textit{d}$_{x^2-y^2}$ of Ti and \textit{p}$_x$ of Se to the \textit{d}$_{x^2-y^2}$ and \textit{d}$_{z^2}$ orbitals of Ti atoms, which accompany the enhancement of electron-phonon coupling in $e$-$h$ recombination process (Figures \ref{fig2},  \ref{fig6}, and S2). Therefore, the averaged NAC values improved from 3.45 meV of FE ground state, 0.26 meV of 0.1 \textit{e}/f.u., and 1.12 meV of 0.25 \textit{e}/f.u. to 49.04 meV of 0.5 \textit{e}/f.u. and 0.8 \textit{e}/f.u. (Figure \ref{fig6}a, c, e, g, and i). The stronger NAC leads to significantly faster $e$-$h$ recombination, reducing the time from 800 to 1 ps. Such significant differences in timescales also support the detection of composite time to determine the occurrence of structural phase transitions, providing another effective detection method for experiments. At high excitation densities (0.5 and 0.8 \textit{e}/f.u.), carrier recombination occurs within $\sim$1 ps, in sharp contrast to the protracted lifetimes (hundreds of ps or longer) observed at lower excitation densities (0.1 and 0.25 \textit{e}/f.u.), reflecting the polarization-assisted carrier separation in these polar materials \cite{wan2023boost,wan2024heterostructures,wanefficient}. Importantly, the ultrafast recombination observed in the highly excited states ensures a prompt recovery to the PE ground state, which is consistent with the reversible light-dark switching demanded for high-performance hydrogen storage.

Our results demonstrate that light irradiation enables dynamic and reversible control of hydrogen-surface bond strength in monolayer ionic ferroelectrics through optically driven ferroic order switching. Using TiGeSe$_3$ as a representative system, we uncover a well-defined photoinduced phase evolution governed by \textit{n}$_\text{ph}$, involving sequential transitions from FE ground state to coexisting FE/S-AFM states (\textit{n}$_\text{ph}$ $>$ 0.2 \textit{e}/f.u.), then to a PE phase with S-AFM order (\textit{n}$_\text{ph}$ $>$ 0.35 \textit{e}/f.u.), and finally to a PE phase with Z-AFM order (\textit{n}$_\text{ph}$ $>$ 0.6 \textit{e}/f.u.). This ferroic transformation substantially reshapes the surface reaction free-energy landscape, increasing $\Delta G_{\text{*H}}$ from 0.33 to 1.11 eV and thereby driving the system from thermoneutral adsorption to spontaneous hydrogen desorption. Importantly, this modulation is continuously tunable via light intensity. NAMD simulations reveal that photoexcitation enhances electron-phonon coupling and nonadiabatic coupling, promoting ultrafast carrier relaxation and facilitating hydrogen desorption through nonthermal pathways. The short \textit{e}-\textit{h} recombination time further ensures rapid restoration of the FE ground state, enabling reversible cycling on picosecond timescales. The correlation between recombination dynamics and ferroic phase evolution also provides a measurable experimental signature for real-time monitoring of light-induced phase transitions. Beyond TiGeSe$_3$, similar light-induced modulation of $\Delta G_{\text{*H}}$ and ferroic switching are validated in AgBiP$_2$Se$_6$ and CuInP$_2$S$_6$ monolayers, highlighting the generality of this mechanism across ionic ferroelectrics. The proposed mechanism can be experimentally validated by correlating pump-probe signatures of ferroic switching with surface-sensitive H$_2$ desorption measurements, providing a clear route to verify the predicted light-controlled adsorption/desorption cycle. Collectively, these findings establish ferroic order as an optically addressable thermodynamic control parameter for surface reactions. This work introduces light-driven ferroic order switching as a general paradigm for dynamically reconfiguring reaction thermodynamics beyond conventional static design strategies.

\subsection*{ACKNOWLEDGMENTS}
This work is supported by the National Natural Science Foundation of China (Grants No.\,12374092, No.\,12374065, and No.\,12374232), the Natural Science Basic Research Program of Shaanxi (Grant No.\,2023-JC-YB-017), the Shaanxi Fundamental Science Research Project for Mathematics and Physics (Grant No.\,22JSQ013), the Open Project of Key Laboratory of Material Simulation Methods \& Software of Ministry of Education, Jilin University (Grant No.\,202604), the “Young Talent Support Plan” of Xi'an Jiaotong University, and the Xiaomi Young Talents Program. C.P. and L.B. thank the Award No. FA9550-23-1-0500 from the U.S. Department of Defense under the DEPSCoR program. L.B. also acknowledges the MonArk NSF Quantum Foundry supported by the National Science Foundation Q-AMASE-i Program under NSF Award No. DMR-1906383 and the Vannevar Bush Faculty Fellowship (VBFF) Grant No. N00014-20-1-2834 from the Department of Defense.

\bibliography{main}

@article{paillard2019photoinduced,
  title={Photoinduced phase transitions in ferroelectrics},
  author={Paillard, Charles and Torun, Engin and Wirtz, Ludger and {\'I}{\~n}iguez, Jorge and Bellaiche, Laurent},
  journal={Phys. Rev. Lett.},
  volume={123},
  number={8},
  pages={087601},
  year={2019},
  publisher={APS},
  doi = {10.1103/PhysRevLett.123.087601},
  url={https://doi.org/10.1103/PhysRevLett.123.087601}
}

@article{paillard2023light,
  title={Light: A new handle to control the structure of cesium lead iodide},
  author={Paillard, Charles and Bellaiche, Laurent},
  journal={Phys. Rev. B},
  volume={107},
  number={5},
  pages={054107},
  year={2023},
  publisher={APS},
  doi = {10.1103/PhysRevB.107.054107},
  url={https://doi.org/10.1103/PhysRevB.107.054107}  
}

@article{lin2014critical,
  title={Critical doping for the onset of a two-band superconducting ground state in $\mathrm{SrTiO_3}$$-$$\delta$},
  author={Lin, Xiao and Bridoux, German and Gourgout, Adrien and Seyfarth, Gabriel and Kr{\"a}mer, Steffen and Nardone, Marc and Fauqu{\'e}, Beno{\^\i}t and Behnia, Kamran},
  journal={Phys. Rev. Lett.},
  volume={112},
  number={20},
  pages={207002},
  year={2014},
  publisher={APS},
  doi = {10.1103/PhysRevLett.112.207002},
  url={https://doi.org/10.1103/PhysRevLett.112.207002}
}

@article{ma2021large,
  title={A large modulation of electron-phonon coupling and an emergent superconducting dome in doped strong ferroelectrics},
  author={Ma, Jiaji and Yang, Ruihan and Chen, Hanghui},
  journal={Nat. Commun.},
  volume={12},
  number={1},
  pages={2314},
  year={2021},
  publisher={Nature Publishing Group UK London},
  doi = {10.1038/s41467-021-22541-1},
  url={https://doi.org/10.1038/s41467-021-22541-1}
}

@article{cazorla2024light,
  title={Light-driven dynamical tuning of the thermal conductivity in ferroelectrics},
  author={Cazorla, Claudio and Bichelmaier, Sebastian and Escorihuela Sayalero, Carlos and {\'I}{\~n}iguez, Jorge and Carrete, Jes{\'u}s and Rurali, Riccardo},
  journal={Nanoscale},
  volume={16},
  number={17},
  pages={8335--8344},
  year={2024},
  publisher={Royal Society of Chemistry},
  doi = {10.1039/D4NR00100A},
  url={https://pubs.rsc.org/en/content/articlehtml/2024/nr/d4nr00100a}
}

@article{Bo2024,
  title = {Photoinduced electronic and spin topological phase transitions in monolayer Bismuth},
  author = {Peng, Bo and Lange, Gunnar F. and Bennett, Daniel and Wang, Kang and Slager, Robert Jan and Monserrat, Bartomeu},
  journal = {Phys. Rev. Lett.},
  volume = {132},
  issue = {11},
  pages = {116601},
  numpages = {8},
  year = {2024},
  month = {Mar},
  publisher = {American Physical Society},
  doi = {10.1103/PhysRevLett.132.116601},
  url = {https://link.aps.org/doi/10.1103/PhysRevLett.132.116601}
}

@article{xu2024electrochemical,
  title={Electrochemical hydrogen storage materials: state-of-the-art and future perspectives},
  author={Xu, Xuelu and Dong, Yue and Hu, Qingwen and Si, Nan and Zhang, Chunwei},
  journal={Energy \& Fuels},
  volume={38},
  number={9},
  pages={7579--7613},
  year={2024},
  publisher={ACS Publications},
  doi = {10.1021/acs.energyfuels.3c05138},
  url={https://pubs.acs.org/doi/full/10.1021/acs.energyfuels.3c05138}
}

@article{hassan2021monetization,
  title={Monetization of the environmental damage caused by fossil fuels},
  author={Hassan, Ather and Ilyas, Syed Zafar and Jalil, Abdul and Ullah, Zahid},
  journal={	Environ. Sci. Pollut. Res.},
  volume={28},
  number={17},
  pages={21204--21211},
  year={2021},
  publisher={Springer},
  doi = {10.1007/s11356-020-12205-w},
  url={https://link.springer.com/article/10.1007/s11356-020-12205-w}
}

@article{majid2020renewable,
  title={Renewable energy for sustainable development in India: current status, future prospects, challenges, employment, and investment opportunities},
  author={Charles Rajesh Kumar. J and M. A. Majid },
  journal={Energy Sustain Soc},
  volume={10},
  number={1},
  pages={1--36},
  year={2020},
  publisher={Springer},
  doi = {10.1186/s13705-019-0232-1},
  url={https://link.springer.com/article/10.1186/s13705-019-0232-1}
}

@article{yadav2022electrochemical,
  title={Electrochemical hydrogen storage: Achievements, emerging trends, and perspectives},
  author={Yadav, Suraj and Oberoi, Amandeep Singh and Mittal, Madhup Kumar},
  journal={Int. J. Energy Res.},
  volume={46},
  number={12},
  pages={16316--16335},
  year={2022},
  publisher={Wiley Online Library},
  doi = {10.1002/er.8407},
  url={https://onlinelibrary.wiley.com/doi/full/10.1002/er.8407}
}

@article{eftekhari2017electrochemical,
  title={Electrochemical hydrogen storage: opportunities for fuel storage, batteries, fuel cells, and supercapacitors},
  author={Eftekhari, Ali and Fang, Baizeng},
  journal={Int. J. Hydrogen Energy},
  volume={42},
  number={40},
  pages={25143--25165},
  year={2017},
  publisher={Elsevier},
  doi = {10.1016/j.ijhydene.2017.08.103},
  url={https://www.sciencedirect.com/science/article/pii/S0360319917333529}
}

@article{zhou2010electric,
  title={Electric field enhanced hydrogen storage on polarizable materials substrates},
  author={Zhou, J and Wang, Q and Sun, Qiang and Jena, Puru and Chen, XS},
  journal={Proc. Natl. Acad. Sci.},
  volume={107},
  number={7},
  pages={2801--2806},
  year={2010},
  publisher={National Academy of Sciences},
  doi = {10.1073/pnas.0905571107},
  url={https://www.pnas.org/doi/abs/10.1073/pnas.0905571107}
}

@article{liu2009electric,
  title={Electric field induced reversible switch in hydrogen storage based on single-layer and bilayer graphenes},
  author={Liu, W and Zhao, Y.H. and Nguyen, J and Li, Y and Jiang, Q and Lavernia, E.J.},
  journal={Carbon},
  volume={47},
  number={15},
  pages={3452--3460},
  year={2009},
  publisher={Elsevier},
  doi = {10.1016/j.carbon.2009.08.012},
  url={https://www.sciencedirect.com/science/article/pii/S0008622309005156}
}

@article{han2024insights,
  title={Insights into the mechanism of electric field regulating hydrogen adsorption on $\mathrm{Li}$-functionalized $\mathrm{N}$-doped defective graphene: A first-principles perspective},
  author={Han, Yong and Ni, Yufeng and Guo, Xiaoqiang and Jiao, Tifeng},
  journal={Fuel},
  volume={357},
  pages={129655},
  year={2024},
  publisher={Elsevier},
  doi = {10.1016/j.fuel.2023.129655},
  url={https://www.sciencedirect.com/science/article/pii/S001623612302269X}
}

@article{zhang2019electric,
  title={Electric-field induced structural transition in vertical $\mathrm{MoTe_2}$- and $\mathrm{Mo_{1-x}W_xTe_2}$- based resistive memories},
  author={Zhang, Feng and Zhang, Huairuo and Krylyuk, Sergiy and Milligan, Cory A and Zhu, Yuqi and Zemlyanov, Dmitry Y and Bendersky, Leonid A and Burton, Benjamin P and Davydov, Albert V and Appenzeller, Joerg},
  journal={Nat. Mater.},
  volume={18},
  number={1},
  pages={55--61},
  year={2019},
  doi = {10.1038/s41563-018-0234-y},
  publisher={Nature Publishing Group UK London},
  url={https://www.nature.com/articles/s41563-018-0234-y}
}

@article{wang2017suppressing,
  title={Suppressing electron--phonon coupling through laser-induced phase transition},
  author={Wang, Zhaowu and Li, Xiyu and Zhang, Guozhen and Luo, Yi and Jiang, Jun},
  journal={ACS Appl. Mater. Interfaces},
  volume={9},
  number={28},
  pages={23309--23313},
  year={2017},
  publisher={ACS Publications},
  doi = {10.1021/acsami.7b05480},
  url={https://pubs.acs.org/doi/full/10.1021/acsami.7b05480}
}

@article{Zhang2019light,
  title = {Light-induced subpicosecond lattice symmetry switch in $\mathrm{MoTe_2}$},
  author = {Zhang, M. Y. and Wang, Z. X. and Li, Y. N. and Shi, L. Y. and Wu, D. and Lin, T. and Zhang, S. J. and Liu, Y. Q. and Liu, Q. M. and Wang, J. and Dong, T. and Wang, N. L.},
  journal = {Phys. Rev. X},
  volume = {9},
  issue = {2},
  pages = {021036},
  numpages = {9},
  year = {2019},
  month = {May},
  publisher = {American Physical Society},
  doi = {10.1103/PhysRevX.9.021036},
  url = {https://link.aps.org/doi/10.1103/PhysRevX.9.021036}
}

@article{sternbach2021nanotextured,
  title={Nanotextured dynamics of a light-induced phase transition in $\mathrm{VO_2}$},
  author={Sternbach, Aaron J and Ruta, Francesco L and Shi, Yin and Slusar, Tetiana and Schalch, Jacob and Duan, Guangwu and McLeod, Alexander S and Zhang, Xin and Liu, Mengkun and Millis, Andrew J and Hyun-Tak Kim and Long-Qing Chen and Richard D. Averitt and D. N. Basov},
  journal={Nano Lett.},
  volume={21},
  number={21},
  pages={9052--9060},
  year={2021},
  publisher={ACS Publications},
  doi = {10.1021/acs.nanolett.1c02638},
  URL={https://pubs.acs.org/doi/full/10.1021/acs.nanolett.1c02638}
}

@article{johnson2023ultrafast,
  title   = {Ultrafast X-ray imaging of the light-induced phase transition in $\mathrm{VO_2}$},
  author  = {Allan S Johnson and Daniel Perez Salinas and Khalid M Siddiqui and Sungwon Kim and Sungwook Choi and Klara Volckaert and Paulina E Majchrzak and S{\o}ren Ulstrup and Naman Agarwal and Kent Hallman and Richard F. Haglund Jr and Christian M. G{\"u}nther and Bastian Pfau and Stefan Eisebitt and Dirk Backes and Francesco Maccherozzi and Ann Fitzpatrick and Sarnjeet S. Dhesi and Pierluigi Gargiani and Manuel Valvidares and Nongnuch Artrith and Frank de Groot and Hyeongi Choi and Dogeun Jang and Abhishek Katoch and Soonnam Kwon and Sang Han Park and Hyunjung Kim and Simon E. Wall},
  journal = {Nat. Phys.},
  volume  = {19},
  number  = {2},
  pages   = {215--220},
  year    = {2023},
  publisher = {Nature Publishing Group UK London},
  doi     = {10.1038/s41567-022-01848-w},
  url     = {https://www.nature.com/articles/s41567-022-01848-w}
}

@article{luchtefeld2023operando,
  title={Operando dissolution of RF magnetron-sputtered thin-film $\mathrm{LoCoO_2}$ cathodes during galvanostatic cycling: In situ evaluation of dissolution and performance data},
  author={Lüchtefeld, Janik and Hemmelmann, Hendrik and Wachs, Susanne and Behling, Christopher and Mayrhofer, Karl JJ and Elm, Matthias T and Berkes, Bal{\'a}zs B},
  journal={J. Phys. Chem. C},
  volume={127},
  number={43},
  pages={21211--21219},
  year={2023},
  publisher={ACS Publications},
  doi = {10.1021/acs.jpcc.3c05587},
  url={https://pubs.acs.org/doi/full/10.1021/acs.jpcc.3c05587}
}

@article{xue20193d,
  title={3D $\mathrm{LoCoO_2}$ nanosheets assembled nanorod arrays via confined dissolution-recrystallization for advanced aqueous lithium-ion batteries},
  author={Xue, Liang and Zhang, Qinghua and Zhu, Xiaohui and Gu, Lin and Yue, Jili and Xia, Qiuying and Xing, Ting and Chen, Tingting and Yao, Yao and Xia, Hui},
  journal={Nano Energy},
  volume={56},
  pages={463--472},
  year={2019},
  publisher={Elsevier},
  doi = {10.1016/j.nanoen.2018.11.085},
  url={https://www.sciencedirect.com/science/article/pii/S2211285518308978}
}

@article{brongersma2015plasmon,
  title={Plasmon-induced hot carrier science and technology},
  author={Brongersma, Mark L and Halas, Naomi J and Nordlander, Peter},
  journal={Nat. Nanotechnol.},
  volume={10},
  number={1},
  pages={25--34},
  year={2015},
  publisher={Nature Publishing Group UK London},
  doi = {10.1038/nnano.2014.311},
  url={https://www.nature.com/articles/nnano.2014.311}
}

@article{sandor2024ultrafast,
  title={Ultrafast surface plasmon probing of interband and intraband hot electron excitations},
  author={S{\'a}ndor, P{\'e}ter and Lov{\'a}sz, B{\'e}la and Budai, Judit and P{\'a}pa, Zsuzsanna and Dombi, P{\'e}ter},
  journal={Nano Lett.},
  volume={24},
  number={26},
  pages={8024--8029},
  year={2024},
  publisher={ACS Publications},
  doi = {10.1021/acs.nanolett.4c01669},
  url={https://pubs.acs.org/doi/full/10.1021/acs.nanolett.4c01669}
}

@article{avdizhiyan2023ultrafast,
  title={Ultrafast laser-induced dynamics of non-equilibrium electron spill-out in nanoplasmonic bilayers},
  author={Avdizhiyan, Artur and Janus, Weronika and Szpytma, Marcin and Slezak, Tomasz and Przybylski, Marek and Chrobak, Maciej and Roddatis, Vladimir and Stupakiewicz, Andrzej and Razdolski, Ilya},
  journal={Nano Lett.},
  volume={24},
  number={1},
  pages={466--471},
  year={2023},
  publisher={ACS Publications},
  doi = {10.1021/acs.nanolett.3c04318},
  url={https://pubs.acs.org/doi/full/10.1021/acs.nanolett.3c04318}
}

@article{bakalis2024momentum,
  title={Momentum-space observation of optically excited nonthermal electrons in graphene with persistent pseudospin polarization},
  author={Bakalis, Jin and Chernov, Sergii and Li, Ziling and Kunin, Alice and Withers, Zachary H and Cheng, Shuyu and Adler, Alexander and Zhao, Peng and Corder, Christopher and White, Michael G and Gerd Schönhense and Xu DuRol and K. Kawakami and Thomas K. Allison},
  journal={Nano Lett.},
  volume={24},
  number={30},
  pages={9353--9359},
  year={2024},
  publisher={ACS Publications},
  doi = {10.1021/acs.nanolett.4c02378},
  url={https://pubs.acs.org/doi/full/10.1021/acs.nanolett.4c02378}
}

@article{shen2025interplay,
  title={Interplay of ultrafast electron--phonon and electron--electron scattering in $\mathrm{Ti_3C_2T_x}$ $\mathrm{MXenes}$: Ab initio quantum dynamics},
  author={Shen, Shiying and Lu, Haoran and Gumber, Shriya and Prezhdo, Oleg V and Long, Run},
  journal={Nano Lett.},
  volume={25},
  number={18},
  pages={7517--7523},
  year={2025},
  publisher={ACS Publications},
  doi = {10.1021/acs.nanolett.5c01242},
  url={https://pubs.acs.org/doi/full/10.1021/acs.nanolett.5c01242}
}

@article{wang2023alterferroicity,
  title={Alterferroicity with seesaw-type magnetoelectricity},
  author={Wang, Ziwen and Dong, Shuai},
  journal={Proc. Natl. Acad. Sci.},
  volume={120},
  number={49},
  pages={e2305197120},
  year={2023},
  publisher={National Academy of Sciences},
  doi = {10.1073/pnas.2305197120},
  url={https://www.pnas.org/doi/abs/10.1073/pnas.2305197120}
}

@article{zhang2025landau,
  title={Landau theory description of autferroicity},
  author={Zhang, Jun-Jie and Yakobson, Boris I and Dong, Shuai},
  journal={Phys. Rev. Lett.},
  volume={134},
  number={21},
  pages={216801},
  year={2025},
  publisher={APS},
  doi = {10.1103/PhysRevLett.134.216801},
  url={https://journals.aps.org/prl/abstract/10.1103/PhysRevLett.134.216801}
}

@article{zheng2019ab,
  title={Ab initio nonadiabatic molecular dynamics investigations on the excited carriers in condensed matter systems},
  author={Zheng, Qijing and Chu, Weibin and Zhao, Chuanyu and Zhang, Lili and Guo, Hongli and Wang, Yanan and Jiang, Xiang and Zhao, Jin},
  journal={Wiley Interdiscip. Rev.:Comput. Mol. Sci.},
  volume={9},
  number={6},
  pages={e1411},
  year={2019},
  publisher={Wiley Online Library},
  doi = {10.1002/wcms.1411},
  url={https://wires.onlinelibrary.wiley.com/doi/full/10.1002/wcms.1411}
}

@article{chu2020low,
  title={Low-frequency lattice phonons in halide perovskites explain high defect tolerance toward electron-hole recombination},
  author={Chu, Weibin and Zheng, Qijing and Prezhdo, Oleg V and Zhao, Jin and Saidi, Wissam A},
  journal={Sci. Adv.},
  volume={6},
  number={7},
  pages={eaaw7453},
  year={2020},
  publisher={American Association for the Advancement of Science},
  doi = {10.1126/sciadv.aaw7453},
  url={https://www.science.org/doi/full/10.1126/sciadv.aaw7453}
}

@article{zheng2017phonon,
  title={Phonon-assisted ultrafast charge transfer at van der Waals heterostructure interface},
  author={Zheng, Qijing and Saidi, Wissam A and Xie, Yu and Lan, Zhenggang and Prezhdo, Oleg V and Petek, Hrvoje and Zhao, Jin},
  journal={Nano. lett.},
  volume={17},
  number={10},
  pages={6435--6442},
  year={2017},
  publisher={ACS Publications},
  doi = {10.1021/acs.nanolett.7b03429},
  url={https://pubs.acs.org/doi/full/10.1021/acs.nanolett.7b03429}
}

@article{yang2019ultrafast,
  title={Ultrafast self-trapping of photoexcited carriers sets the upper limit on antimony trisulfide photovoltaic devices},
  author={Yang, Zhaoliang and Wang, Xiaomin and Chen, Yuzhong and Zheng, Zhenfa and Chen, Zeng and Xu, Wenqi and Liu, Weimin and Yang, Yang and Zhao, Jin and Chen, Tao and  Haiming Zhu },
  journal={Nat. Commun.},
  volume={10},
  number={1},
  pages={4540},
  year={2019},
  publisher={Nature Publishing Group UK London},
  doi = {10.1038/s41467-019-12445-6},
  url={https://www.nature.com/articles/s41467-019-12445-6}
}

@article{yao2022density,
  title={Density functional theory study on the enhancement mechanism of the photocatalytic properties of the g-$\mathrm{C_3N_4}$/$\mathrm{BiOBr}$(001) heterostructure},
  author={Yao, Wenzhi and Li, Dongying and Wei, Shuai and Liu, Xiaoqing and Liu, Xuefei and Wang, Wentao},
  journal={ACS omega},
  volume={7},
  number={41},
  pages={36479--36488},
  year={2022},
  publisher={ACS Publications},
  doi = {10.1021/acsomega.2c04298},
  url={https://pubs.acs.org/doi/full/10.1021/acsomega.2c04298}
}

@article{ge2023two,
  title={Two-dimensional $\mathrm{PtI_2/Bi_2S_3}$ and $\mathrm{PtI_2/Bi_2Se_3}$ heterostructures with high solar-to-hydrogen efficiency},
  author={Ge, Meng and Yang, Chuan Lu and Wang, Mei Shan and Ma, Xiao Guang},
  journal={Colloids Surf., A},
  volume={666},
  pages={131286},
  year={2023},
  publisher={Elsevier},
  doi = {10.1016/j.colsurfa.2023.131286},
  url={https://www.sciencedirect.com/science/article/pii/S0927775723003709}
}

@article{sun2022high,
  title={High solar-to-hydrogen efficiency photocatalytic hydrogen evolution reaction with the $\mathrm{HfSe_2/InSe}$ heterostructure},
  author={Sun, Rui and Yang, Chuan Lu and Wang, Mei Shan and Ma, Xiao Guang},
  journal={J. Power Sources},
  volume={547},
  pages={232008},
  year={2022},
  publisher={Elsevier},
  doi = {10.1016/j.jpowsour.2022.232008},
  url={https://www.sciencedirect.com/science/article/pii/S0378775322009855}
}

@article{fu2018intrinsic,
  title={Intrinsic electric fields in two-dimensional materials boost the solar-to-hydrogen efficiency for photocatalytic water splitting},
  author={Fu, Cen Feng and Sun, Jiuyu and Luo, Qiquan and Li, Xingxing and Hu, Wei and Yang, Jinlong},
  journal={Nano Lett.},
  volume={18},
  number={10},
  pages={6312--6317},
  year={2018},
  publisher={ACS Publications},
  doi = {10.1021/acs.nanolett.8b02561#_i9},
  url={https://pubs.acs.org/doi/10.1021/acs.nanolett.8b02561#_i9}
}

@article{wan2023boost,
  title={Boost solar-to-hydrogen efficiency by constructing heterostructures with the pristine and $\mathrm{Se/Te}$-doped $\mathrm{AgInP_2S_6}$ monolayers},
  author={Wan, Xue-Qing and Yang, Chuan-Lu and Wang, Mei-Shan and Ma, Xiao-Guang},
  journal={Appl. Surf. Sci.},
  volume={614},
  pages={156254},
  year={2023},
  publisher={Elsevier},
  doi = {10.1016/j.apsusc.2022.156254},
  url={https://www.sciencedirect.com/science/article/pii/S0169433222037825}
}

@article{zhu2021intrinsic,
  title={Intrinsic $\mathrm{ORR}$ activity enhancement of $\mathrm{Pt}$ atomic sites by engineering the d-band center via local coordination tuning},
  author={Zhu, Xiaofeng and Tan, Xin and Wu, Kuang Hsu and Haw, Shu Chih and Pao, Chih Wen and Su, Bing Jian and Jiang, Junjie and Smith, Sean C and Chen, Jin Ming and Amal, Rose and  Xunyu Lu},
  journal={Angew. Chem., Int. Ed.},
  volume={60},
  number={40},
  pages={21911--21917},
  year={2021},
  publisher={Wiley Online Library},
  doi = {10.1002/anie.202107790},
  url={https://onlinelibrary.wiley.com/doi/full/10.1002/anie.202107790}
}

@article{jiao2022descriptors,
  title={Descriptors for the evaluation of electrocatalytic reactions: d-band theory and beyond},
  author={Jiao, Shilong and Fu, Xianwei and Huang, Hongwen},
  journal={Adv. Funct. Mater.},
  volume={32},
  number={4},
  pages={2107651},
  year={2022},
  publisher={Wiley Online Library},
  doi = {10.1002/adfm.202107651},
  url={https://advanced.onlinelibrary.wiley.com/doi/full/10.1002/adfm.202107651}
}

@article{norskov2011density,
  title={Density functional theory in surface chemistry and catalysis},
  author={N{\o}rskov, Jens K and Abild-Pedersen, Frank and Studt, Felix and Bligaard, Thomas},
  journal={Proc. Natl. Acad. Sci.},
  volume={108},
  number={3},
  pages={937--943},
  year={2011},
  publisher={National Academy of Sciences},
  doi = {10.1073/pnas.1006652108},
 url={https://www.pnas.org/doi/abs/10.1073/pnas.1006652108}
}

@article{zhang2023engineering,
  title={Engineering p-band center of oxygen boosting $\mathrm{H}$$^+$ intercalation in $\delta$-$\mathrm{MnO}$$_2$ for aqueous zinc ion batteries},
  author={Zhang, Jianhua and Li, Wenbin and Wang, Jingjing and Pu, Xiaohua and Zhang, Gaini and Wang, Shuai and Wang, Ni and Li, Xifei},
  journal={Angew. Chem.},
  volume={135},
  number={8},
  pages={e202215654},
  year={2023},
  publisher={Wiley Online Library},
  doi = {10.1002/ange.202215654},
  url={https://onlinelibrary.wiley.com/doi/full/10.1002/ange.202215654}
}

@article{zhang2025unraveling,
  title={Unraveling subthermionic transport in one-dimensional van der Waals isolated-band $\mathrm{FETs}$},
  author={Zhang, Weiming and Li, Kaiqi and Wang, Bing and Sun, Yuqi and Zhou, Jian and Sun, Zhimei},
  journal={J. Phys. Chem. Lett.},
  volume={16},
  number={19},
  pages={4698--4706},
  year={2025},
  publisher={ACS Publications},
  doi = {10.1021/acs.jpclett.5c00824},
  url={https://pubs.acs.org/doi/full/10.1021/acs.jpclett.5c00824}
}

@article{yu2010conduction,
  title={Conduction band energy level control of titanium dioxide: toward an efficient visible-light-sensitive photocatalyst},
  author={Yu, Huogen and Irie, Hiroshi and Hashimoto, Kazuhito},
  journal={J. Am. Chem. Soc.},
  volume={132},
  number={20},
  pages={6898--6899},
  year={2010},
  publisher={ACS Publications},
  doi = {10.1021/ja101714s},
  url={https://pubs.acs.org/doi/full/10.1021/ja101714s}
}

@article{irie2007ag+,
  title={$\mathrm{Ag}$$^+$- and $\mathrm{Pb}$$^{2+}$-doped $\mathrm{SrTiO_3}$ photocatalysts. A correlation between band structure and photocatalytic activity},
  author={Irie, Hiroshi and Maruyama, Yoshihiko and Hashimoto, Kazuhito},
  journal={J. Phys. Chem. C},
  volume={111},
  number={4},
  pages={1847--1852},
  year={2007},
  publisher={ACS Publications},
  doi = {10.1021/jp066591i},
  url={https://pubs.acs.org/doi/full/10.1021/jp066591i}
}

@article{wanefficient,
  title={Efficient Z-scheme photocatalyst for Hydrogen production via water splitting using $\mathrm{CH_3}$-and $\mathrm{F}$-modified $\mathrm{C_{60}}$ Fullerene-based heterostructures},
  author={Wan, Xue-Qing and Yang, Chuan-Lu and Shi, Wen-Jie and Li, Xiaohu and Liu, Yuliang and Zhao, Wenkai and Gao, Feng},
  journal={Small},
  volume={22},
  pages={2504146},
  year={2025},
  publisher={Wiley Online Library},
  doi = {10.1002/smll.202504146},
  url={https://onlinelibrary.wiley.com/doi/full/10.1002/smll.202504146}
}

@article{wan2024heterostructures,
  title={Heterostructures stacked with $\mathrm{X_2SY}$ ($\mathrm{X}$= $\mathrm{In}$, $\mathrm{Ga}$; $\mathrm{Y}$= $\mathrm{Se}$, $\mathrm{Te}$) and g-$\mathrm{C_2N}$ monolayers for high power conversion efficiency solar cells: insight from electronic properties and nonadiabatic dynamics},
  author={Wan, Xue-Qing and Yang, Chuan-Lu and Li, Xiao-Hu and Liu, Yu-Liang and Zhao, Wen-Kai},
  journal={J. Mater. Chem. A},
  volume={12},
  number={27},
  pages={16559--16568},
  year={2024},
  publisher={Royal Society of Chemistry},
  doi = {10.1039/D4TA01263A},
  url={https://pubs.rsc.org/en/content/articlehtml/2024/ta/d4ta01263a}
}

@article{chen2024unusual,
  title={Unusual Sabatier principle on high entropy alloy catalysts for hydrogen evolution reactions},
  author={Chen, Zhi Wen and Li, Jian and Ou, Pengfei and Huang, Jianan Erick and Wen, Zi and Chen, LiXin and Yao, Xue and Cai, GuangMing and Yang, Chun Cheng and Singh, Chandra Veer and Qing Jiang},
  journal={Nat. Commun.},
  volume={15},
  number={1},
  pages={359},
  year={2024},
  publisher={Nature Publishing Group UK London},
  doi = {10.1038/s41467-023-44261-4},
  url={https://www.nature.com/articles/s41467-023-44261-4}
}

@book{sabatier1920catalyse,
  title = {La catalyse en chimie organique},
  author = {Sabatier, P.},
  publisher = {C. B{\'e}ranger},
  address = {Paris},
  year = {1920},
  series = {Encyclop{\'e}die de Science Chimique Appliqu{\'e}e aux Arts Industriels},
  volume = {},
  pages = {},
  doi = {},
  url = {https://books.google.com.hk/books?id=pIcMAQAAIAAJ}
}

@article{jaeger2012decoherence,
  title={Decoherence-induced surface hopping},
  author={Jaeger, Heather M and Fischer, Sean and Prezhdo, Oleg V},
  journal={J. Chem. Phys.},
  volume={137},
  number={22},
  pages = {22A545},
  year={2012},
  publisher={AIP Publishing},
  doi = {10.1063/1.4757100},
  url={https://pubs.aip.org/aip/jcp/article/137/22/22A545/194968}
}

@article{king1993,
  title = {Theory of polarization of crystalline solids},
  author = {King-Smith, R. D. and Vanderbilt, David},
  journal = {Phys. Rev. B},
  volume = {47},
  issue = {3},
  pages = {1651--1654},
  numpages = {0},
  year = {1993},
  month = {Jan},
  publisher = {American Physical Society},
  doi = {10.1103/PhysRevB.47.1651},
  url = {https://link.aps.org/doi/10.1103/PhysRevB.47.1651}
}

@article{long2016quantum,
  title={Quantum coherence facilitates efficient charge separation at a $\mathrm{MoS_2}$/$\mathrm{MoSe_2}$ van der Waals junction},
  author={Long, Run and Prezhdo, Oleg V},
  journal={Nano Lett.},
  volume={16},
  number={3},
  pages={1996--2003},
  year={2016},
  publisher={ACS Publications},
  doi = {10.1021/acs.nanolett.5b05264},
  url={https://pubs.acs.org/doi/abs/10.1021/acs.nanolett.5b05264}
}

@article{sun2019strain,
  title={Strain engineering to facilitate the occurrence of $\mathrm{2D}$ ferroelectricity in $\mathrm{CuInP_2S_6}$ monolayer},
  author={Sun, Zhi-Zheng and Xun, Wei and Jiang, Li and Zhong, Jia-Lin and Wu, Yin-Zhong},
  journal={J. Phys. D:Appl. Phys.},
  volume={52},
  number={46},
  pages={465302},
  year={2019},
  publisher={IOP Publishing},
  doi = {10.1088/1361-6463/ab3aa7/meta},
  url={https://iopscience.iop.org/article/10.1088/1361-6463/ab3aa7/meta}
}

@article{qi2018two,
  title={Two-dimensional multiferroic semiconductors with coexisting ferroelectricity and ferromagnetism},
  author={Qi, Jingshan and Wang, Hua and Chen, Xiaofang and Qian, Xiaofeng},
  journal={Appl. Phys. Lett.},
  volume={113},
  pages={043102},
  number={4},
  year={2018},
  publisher={AIP Publishing},
  doi = {10.1063/1.5038037},
  url={https://pubs.aip.org/aip/apl/article/113/4/043102/286499}
}

@article{kim2024interfacial,
  title={Interfacial charge transfer driven by surface termination-controlled $\mathrm{Ti_2C}$ $\mathrm{MXene}$ for enhanced hydrogen storage in magnesium},
  author={Kim, Min Gyu and Kang, ShinYoung and Wood, Brandon C and Cho, Eun Seon},
  journal={ J. Mater. Chem. A},
  volume={12},
  number={40},
  pages={27212--27219},
  year={2024},
  publisher={Royal Society of Chemistry},
 doi = {10.1039/D4TA04563G},
  url={https://pubs.rsc.org/en/content/articlehtml/2024/ta/d4ta04563g}
}

@article{liu2024mg,
  title={$\mathrm{Mg}$-$\mathrm{MOF}$-74 derived defective framework for hydrogen storage at above-ambient temperature assisted by $\mathrm{Pt}$ catalyst},
  author={Liu, Shiyuan and Zhang, Yue and Zhu, Fangzhou and Liu, Jieyuan and Wan, Xin and Liu, Ruonan and Liu, Xiaofang and Shang, Jiaxiang and Yu, Ronghai and Feng, Qiang and  Zili, Wang and Jianglan, Shui},
  journal={Adv. Sci.},
  volume={11},
  number={18},
  pages={2401868},
  year={2024},
  publisher={Wiley Online Library},
  doi = {10.1002/advs.202401868},
  url={https://advanced.onlinelibrary.wiley.com/doi/full/10.1002/advs.202401868}
}

@article{parkar2024hydrogen,
  title={Hydrogen storage properties of $\mathrm{Ti}$-doped $\mathrm{C_{20}}$ nanocage and its derivatives: A comprehensive density functional theory investigation},
  author={Parkar, Poonam and Chaudhari, Ajay},
  journal={	Mater. Chem. Phys.},
  volume={319},
  pages={129340},
  year={2024},
  publisher={Elsevier},
  doi = {10.1016/j.matchemphys.2024.129340},
  url={https://www.sciencedirect.com/science/article/pii/S0254058424004656}
}

@article{gao2025rationally,
  title={Rationally designed carbon nanomaterials for electrically driven solid-state hydrogen storage},
  author={Gao, Yong and Gao, Panyu and Li, Chao and Yue, Qiuyan and Liang, Qing and Qiao, Sifan and Zhang, Wei and Zheng, Weitao and Zhang, Lipeng and Li, Zhenglong and Cui, Wen-Gang and Wang, Xiaowei and Wan, Yiyang and Zhang, Mingchang and Wang, Xinqiang and Liu, Yanxia and Qi, Fulai and Li, Chenchen and Miao, Jian and Zhang, Jing and Han, Xiao and Wang, Pan and Guo, Chang and Chen, Qiao and Xu, Ziyuan and Gao, Mingxia and Sun, Wenping and Yang, Yaxiong and Chen, Jian and Xia, Zhenhai and Pan, Hongge},
  journal={Adv. Funct. Mater.},
  pages={e05188},
  volume = {35},
  number = {49},
  year={2025},
  publisher={Wiley Online Library},
  doi = {10.1002/adfm.202505188},
  url={https://advanced.onlinelibrary.wiley.com/doi/full/10.1002/adfm.202505188}
}

@article{verma2024study,
  title={Study of dual osmium and boron co-doped $\mathrm{SWCNTs}$ for reversible hydrogen storage},
  author={Verma, Ritu and Jaggi, Neena},
  journal={Diamond Relat. Mater.},
  volume={148},
  pages={111470},
  year={2024},
  publisher={Elsevier},
  doi = {10.1016/j.diamond.2024.111470},
  url={https://www.sciencedirect.com/science/article/pii/S0925963524006836}
}

@article{qiao2023effects,
  title={Effects of $\mathrm{Cu}$ doping on the hydrogen storage performance of $\mathrm{Ti}$-$\mathrm{Mn}$-based, $\mathrm{AB_2}$-type alloys},
  author={Qiao, Wenfeng and Yin, Dongming and Zhao, Shaolei and Ding, Nan and Liang, Long and Wang, Chunli and Wang, Limin and He, Miao and Cheng, Yong},
  journal={Chem. Eng. J.},
  volume={465},
  pages={142837},
  year={2023},
  publisher={Elsevier},
  doi = {10.1016/j.cej.2023.142837},
  url={https://www.sciencedirect.com/science/article/pii/S1385894723015681}
}

@article{seif2016new,
  title={A new strategy for hydrogen storage using $\mathrm{BNNS}$: simultaneous effects of doping and charge modulation},
  author={Seif, Abdolvahab and Azizi, Khaled},
  journal={RSC Adv.},
  volume={6},
  number={63},
  pages={58458--58468},
  year={2016},
  publisher={Royal Society of Chemistry},
  doi = {10.1039/c6ra06634h },
  url={https://pubs.rsc.org/en/content/articlehtml/2016/ra/c6ra06634h?casa_token=6OWXq3J2FBEAAAAA:3wnE4VwYfPo7ggOUoKRousDPwLONbR73rjiRqskzDTf6LU8uUjhj1SGBuYwdPWbMldG6adJo2lbLDg&casa_token=u0vZ7f4F5sUAAAAA:0VS_Ugp34y9OuIxFa82ZxgXV5D6b1zfXRQmatPq6BHhYsw6ua2GBT4WSr614ilMuy2PesGtoXFbVRg}
}

@article{shi2017phosphorus,
  title={Phosphorus-$\mathrm{Mo_2C}$@carbon nanowires toward efficient electrochemical hydrogen evolution: composition, structural and electronic regulation},
  author={Shi, Zhangping and Nie, Kaiqi and Shao, Zheng Jiang and Gao, Boxu and Lin, Huanlei and Zhang, Hongbin and Liu, Bolun and Wang, Yangxia and Zhang, Yahong and Sun, Xuhui and Cao, Xiaoming Hu, P and Gao, Qingsheng and Tang, Yi},
  journal={	Energy Environ. Sci.},
  volume={10},
  number={5},
  pages={1262--1271},
  year={2017},
  publisher={Royal Society of Chemistry},
  doi = {10.1039/C7EE00388A},
  url={https://pubs.rsc.org/en/content/articlehtml/2015/ee/c7ee00388a}
}

@article{xia2021high,
  title={High-density defects activating Fe-doped molybdenum sulfide@$\mathrm{N}$-doped carbon heterostructures for efficient electrochemical hydrogen evolution},
  author={Xia, Xiaohong and Zhao, Gaiyun and Yan, Qi and Wang, Biao and Wang, Qiufeng and Xie, Haijiao},
  journal={ACS Sustainable Chem. Eng.},
  volume={10},
  number={1},
  pages={182--193},
  year={2021},
  publisher={ACS Publications},
  doi = {10.1021/acssuschemeng.1c05538},
  url={https://pubs.acs.org/doi/full/10.1021/acssuschemeng.1c05538}
}

@article{baaddi2024effect,
  title={The effect of strain on hydrogen storage characteristics in $\mathrm{K_2NaAlH_6}$ double perovskite hydride through first principle method},
  author={Baaddi, Malika and Chami, Rachid and Baalla, Oumaima and Quaoubi, Soukaina El and Saadi, Ali and Omari, Lhaj El Hachemi and Chafi, Mohammed},
  journal={Environ. Sci. Pollut. Res.},
  volume={31},
  number={53},
  pages={62056--62064},
  year={2024},
  publisher={Springer},
  doi = {10.1007/s11356-023-27529-6},
  url={https://link.springer.com/article/10.1007/s11356-023-27529-6}
}

@article{li2020single,
  title={Single-metal atoms supported on $\mathrm{MBenes}$ for robust electrochemical hydrogen evolution},
  author={Li, Bing and Wu, Yang and Li, Neng and Chen, Xingzhu and Zeng, Xianbing and Arramel and Zhao, Xiujian and Jiang, Jizhou},
  journal={ACS Appl. Mater. Interfaces},
  volume={12},
  number={8},
  pages={9261--9267},
  year={2020},
  publisher={ACS Publications},
  doi = {10.1021/acsami.9b20552},
  url={https://pubs.acs.org/doi/full/10.1021/acsami.9b20552}
}

@article{zhou2017improvement,
  title={Improvement in low-temperature and instantaneous high-rate output performance of $\mathrm{Al}$-free $\mathrm{AB_5}$-type hydrogen storage alloy for negative electrode in $\mathrm{Ni/MH}$ battery: effect of thermodynamic and kinetic regulation via partial $\mathrm{Mn}$ substituting},
  author={Zhou, Wanhai and Zhu, Ding and Tang, Zhengyao and Wu, Chaoling and Huang, Liwu and Ma, Zhewen and Chen, Yungui},
  journal={J. Power Sources},
  volume={343},
  pages={11--21},
  year={2017},
  publisher={Elsevier},
  doi = {10.1016/j.jpowsour.2017.01.023},
  url={https://www.sciencedirect.com/science/article/pii/S0378775317300241}
}

@article{fan2021highly,
  title={Highly efficient photocatalytic $\mathrm{CO_2}$ reduction in two-dimensional ferroelectric $\mathrm{CuInP_2S_6}$ bilayers},
  author={Fan, Yingcai and Song, Xiaohan and Ai, Haoqiang and Li, Weifeng and Zhao, Mingwen},
  journal={ACS Appl. Mater. Interfaces},
  volume={13},
  number={29},
  pages={34486--34494},
  year={2021},
  publisher={ACS Publications},
  doi = {10.1021/acsami.1c10983},
  url={https://pubs.acs.org/doi/full/10.1021/acsami.1c10983}
}

@article{liu2022excited,
  title={Excited-state properties of $\mathrm{CuInP_2S_6}$ monolayer as photocatalyst for water splitting},
  author={Liu, Hongling and Yu, Shiqiang and Wang, Yuanyuan and Huang, Baibiao and Dai, Ying and Wei, Wei},
  journal={J. Phys. Chem. Lett.},
  volume={13},
  number={8},
  pages={1972--1978},
  year={2022},
  publisher={ACS Publications},
  doi = {10.1021/acs.jpclett.2c00105},
  url={https://pubs.acs.org/doi/10.1021/acs.jpclett.2c00105}
}

@article{xu2017monolayer,
  title={Monolayer $\mathrm{AgBiP_2Se_6}$: an atomically thin ferroelectric semiconductor with out-plane polarization},
  author={Xu, Bo and Xiang, Hui and Xia, Yidong and Jiang, Kun and Wan, Xiangang and He, Jun and Yin, Jiang and Liu, Zhiguo},
  journal={Nanoscale},
  volume={9},
  number={24},
  pages={8427--8434},
  year={2017},
  publisher={Royal Society of Chemistry},
  doi = {10.1039/c7nr02461d},
  url={https://pubs.rsc.org/en/content/articlelanding/2017/nr/c7nr02461d}
}

@article{ju2019tunable,
  title={Tunable photocatalytic water splitting by the ferroelectric switch in a 2D $\mathrm{AgBiP_2Se_6}$ monolayer},
  author={Ju, Lin and Shang, Jing and Tang, Xiao and Kou, Liangzhi},
  journal={J. Am. Chem. Soc.},
  volume={142},
  number={3},
  pages={1492--1500},
  year={2019},
  publisher={ACS Publications},
 doi = {10.1021/jacs.9b11614},
  url={https://pubs.acs.org/doi/pdf/10.1021/jacs.9b11614?ref=article_openPDF}
}

@article{gueymard2002proposed,
  title={Proposed reference irradiance spectra for solar energy systems testing},
  author={Gueymard, Christian A and Myers, David and Emery, Keith},
  journal={Solar energy},
  volume={73},
  number={6},
  pages={443--467},
  year={2002},
  publisher={Elsevier},
  url={https://www.sciencedirect.com/science/article/pii/S0038092X03000057}
}

@article{lin2020ferroelectric,
  title={Ferroelectric-field accelerated charge transfer in $\mathrm{2D}$ $\mathrm{CuInP_2S_6}$ heterostructure for enhanced photocatalytic $\mathrm{H_2}$ evolution},
  author={Lin, Bo and Chaturvedi, Apoorva and Di, Jun and You, Lu and Lai, Chen and Duan, Ruihuan and Zhou, Jiadong and Xu, Baorong and Chen, Zihao and Song, Pin and Peng, Juan and Ma, Bowen and Liu,  
  Haishi and Meng, Peng and Yang, Guidong and Zhang, Hua  and Liu, Zheng and Liu, Fucai},
  journal={Nano Energy},
  volume={76},
  pages={104972},
  year={2020},
  publisher={Elsevier},
  url={https://www.sciencedirect.com/science/article/pii/S2211285520305498}
}

@article{yu2021few,
  title={Few-layered $\mathrm{CuInP_2S_6}$ nanosheet with sulfur vacancy boosting photocatalytic hydrogen evolution},
  author={Yu, Peng and Wang, Fengmei and Meng, Jun and Shifa, Tofik Ahmed and Sendeku, Marshet Getaye and Fang, Ju and Li, Shuxian and Cheng, Zhongzhou and Lou, Xiaoding and He, Jun},
  journal={CrystEngComm},
  volume={23},
  number={3},
  pages={591--598},
  year={2021},
  publisher={Royal Society of Chemistry},
  url={http://dx.doi.org/10.1039/D0CE01487G}
}

@article{chiang2024manipulating,
  title={Manipulating ferroelectric polarization and spin polarization of $\mathrm{2D}$ $\mathrm{CuInP_2S_6}$ crystals for photocatalytic $\mathrm{CO_2}$ reduction},
  author={Chiang, Chun-Hao and Lin, Cheng-Chieh and Lin, Yin-Cheng and Huang, Chih-Ying and Lin, Cheng-Han and Chen, Ying-Jun and Ko, Ting-Rong and Wu, Heng-Liang and Tzeng, Wen-Yen and Ho, Sheng-Zhu and Chen, Yi-Chun and Ho, Ching-Hwa and Yang, Cheng-Jie and Cyue, Zih-Wei and Dong, Chung-Li and Luo, Chih-Wei and Chen, Chia-Chun and Chen, Chun-Wei},
  journal={J. Am. Chem. Soc.},
  volume={146},
  number={33},
  pages={23278--23288},
  year={2024},
  publisher={ACS Publications},
  url={https://pubs.acs.org/doi/abs/10.1021/jacs.4c05798}
}

\newpage      

\newpage  

\newpage  

\newpage  

\newpage  

\newpage

\end{document}